\documentclass[twocolumn]{aastex6}

\usepackage{hyperref}
\hypersetup{urlcolor=blue}

\begin{document}


\title{Quasi-Periodic Pulsations during the Impulsive and Decay phases of an X-class Flare}

%% Use \author, \affil, plus the \and command to format author and affiliation 
%% information.  If done correctly the peer review system will be able to
%% automatically put the author and affiliation information from the manuscript
%% and save the corresponding author the trouble of entering it by hand.
%%
%% The \affil should be used to document primary affiliations and the
%% \altaffil should be used for secondary affiliations, titles, or email.

%% Authors with the same affiliation can be grouped in a single
\author{L. A. Hayes\altaffilmark{1,2,3}, P. T. Gallagher\altaffilmark{1}, B. R. Dennis\altaffilmark{2}, J. Ireland\altaffilmark{2,3}, A. R. Inglis\altaffilmark{2,4}, D. F. Ryan\altaffilmark{2,5}}

\altaffiltext{1}{School of Physics, Trinity College Dublin, Dublin 2, Ireland}
\altaffiltext{2}{Solar Physics Laboratory, Heliophysics Science Division, NASA Goddard Space Flight Center, Greenbelt, MD 20771, USA}
\altaffiltext{3}{ADNET Systems, Inc.}
\altaffiltext{4}{Physics Department, The Catholic University of America, Washington, DC, 20664, USA}
\altaffiltext{5}{Solar-Terrestrial Centre for Excellence, Royal Observatory of Belgium, Uccle 1180, Brussels, Belgium}

\begin{abstract}

Quasi-periodic pulsations (QPP) are often observed in X-ray emission from solar flares. To date, it is unclear what their physical origins are. Here, we present a multi-instrument investigation of the nature of QPP during the impulsive and decay phases of the X1.0 flare of 28 October 2013. We focus on the character of the fine structure pulsations evident in the soft X-ray time derivatives and compare this variability with structure across multiple wavelengths including hard X-ray and microwave emission. We find that during the impulsive phase of the flare, high correlations between pulsations in the thermal and non-thermal emissions are seen. A characteristic timescale of $\sim$20~s is observed in all channels and a second timescale of $\sim$55~s is observed in the non-thermal emissions. Soft X-ray pulsations are seen to persist into the decay phase of this flare, up to 20 minutes after the non-thermal emission has ceased. We find that these decay phase thermal pulsations have very small amplitude and show an increase in characteristic timescale from $\sim$40~s up to $\sim$70~s. We interpret the bursty nature of the co-existing multi-wavelength QPP during the impulsive phase in terms of episodic particle acceleration and plasma heating. The persistent thermal decay phase QPP are most likely connected with compressive MHD processes in the post-flare loops such as the fast sausage mode or the vertical kink mode. 
\end{abstract}


\section{Introduction} \label{sec:intro}

During a solar flare, the X-ray flux from the Sun can increase by several orders of magnitude and be accompanied by pulsations in the flare emission. These pulsations, known as quasi-periodic pulsations (QPP), are variations of flux as a function of time. Their nature has been examined in many previous studies of both solar \citep[e.g.][]{parks,kane,asai,fleisman, nak09, rez, li_ning} and stellar \citep[e.g.][]{balona, pugh} observations. In a typical event, the emission from a solar flare is seen to pulsate with a characteristic timescale ranging from $\leqslant$ 1~s up to several minutes. QPP are commonly observed during the impulsive phase of solar flares and have been reported in a wide range of wavelengths from radio and microwaves to hard X-rays (HXR) and $\gamma$-rays. 

Two main interpretations outlined in a recent review by \citet{nak09} have been pursued in order to understand QPP. These are that the observed flux variations are driven either by magnetohydrodynamic (MHD) wave behaviour in the corona, or by periodic or `bursty' energy releases from the coronal magnetic field. 

MHD oscillations are known to be supported in corona \citep{roberts82, roberts83}, and flare generated MHD oscillations have been observationally identified in coronal loops \citep{asch_99, nak_intro}. Various MHD wave modes can alter physical plasma parameters and produce the quasi-periodic behavior in flaring lightcurves \citep{nak09}. These wave modes can modulate the emission directly, affect the dynamics of charged particles, or periodically trigger magnetic reconnection \citep{nak_zim}.
However it remains a challenge to interpret QPP purely in terms of linear MHD oscillation theory given the large modulation depths observed in lightcurves, along with the geometrical evolution that occurs during the impulsive phase of flares.
%Observed periods of some QPP are consistent with expected characteristic periods derived from MHD wave theory \citep[e.g.][]{mcewan, pascoe07, pascoe09, macnamara}, further supporting that QPP may be a signature of MHD oscillations.

 \begin{figure*}
\begin{center}
  \includegraphics[width=1.\linewidth]{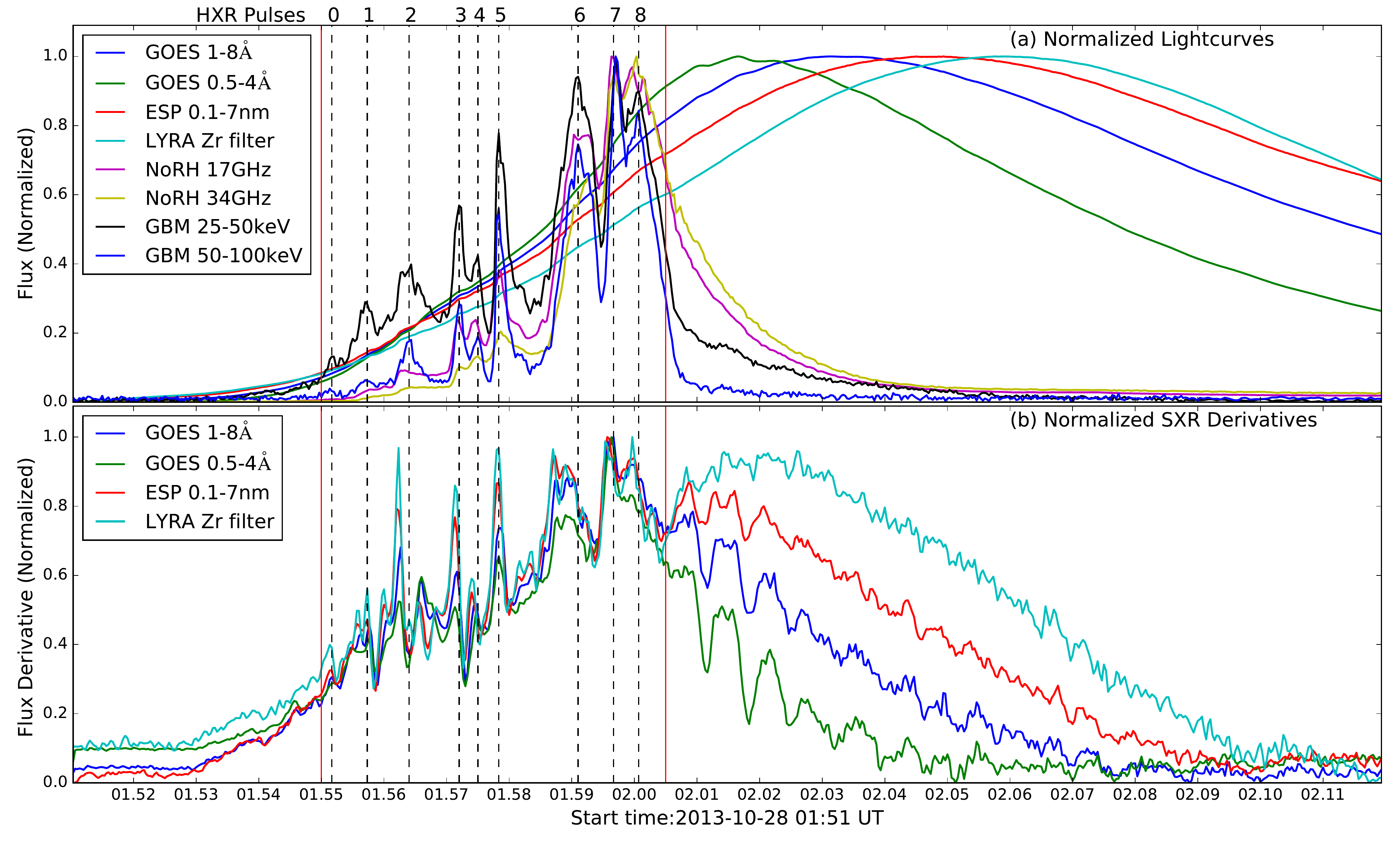}
\caption{(a) Normalized light curves from different instruments for the flare of 2013 October 28. Detector NaI 6 was used for GBM. (b) Derivatives of the soft X-ray channels. The vertical red lines show the start and end of the impulsive phase and the dashed lines show the timing of the HXR pulses.}
\label{fig:one}
\end{center}
\end{figure*}

A second interpretation of QPP is that the pulsations are a direct result of `bursty' regimes of energy release, in particular magnetic reconnection. Early magnetotail studies suggested that magnetic reconnection can happen in an episodic fashion \citep{coppi, sch_coppi}, with subsequent observations supporting this view \citep{hones}. In coronal conditions, recent numerical models have shown that magnetic reconnection can occur in a repetitive regime \citep[e.g.][]{kliem, drake06, linton, guidoni}. Repeated episodes of magnetic reconnection itself could account for the modulation of emission in many different wavelengths. It could explain QPP observed in the non-thermal emission due to changing particle acceleration rates. Variations at other wavelengths such as soft X-ray (SXR) and EUV would then be explained by fluctuations in plasma heating. 

%Present observations do not allow for a choice between these possible theories for flaring pulsations, and it is quite likely that independent cases operate differently or by several non-exclusive processes. In any case, these oscillatory pulsations provide insight into the physical processes and energy mechanisms occurring in flares. Without a complete description of the nature and origin of these pulsations, a full understanding of solar flares is not possible.

The majority of previous flare QPP studies have focused on pulsations observed in emission associated with non-thermal electrons such as HXR and microwave observations. This is due to the large modulation depths observed in this type of emission, especially in the impulsive phase. Recently, however, it has been shown that fine structure pulsations are also evident in the SXR emission \citep{dolla, simoes, dennis}. The nature of these pulsations in thermal emission remains to be studied in detail, and comparisons across multiple wavelengths are required to improve our understanding of the QPP phenomenon.

In this letter, we investigate the nature of these X-ray pulsations in a multi-instrument analysis of a GOES X-class flare, paying particular attention to the fine structure observed during the impulsive and decay phases.

\section{Observations} \label{sec:obs}

Our investigation focuses on pulsations observed in the GOES X1.0 flare on 28 October 2013 (SOL2013-10-28). Examining ten wavelength bands on five different instruments, our data are from high cadence ($\leq$2~s) observations of X-rays (both hard and soft), and microwave emissions from both ground-based and space-borne instruments. Pulsations in the thermal emission were observed using both channels (1--8~\AA\ and 0.5--4~\AA) of the X-ray Sensor on board the \textit{Geostationary Operational Environment Satellite} (GOES 15), the Zirconium channel ($<$2~nm + 6--20~nm) of the Large Yield Radiometer \citep[LYRA;][]{dom} on the \textit{Project for On-Board Autonomy} (PROBA2), and the SXR channel (0.1--7~nm) of the Extreme Ultraviolet Spectrophotometer (ESP), which is part of the EUV Experiment \citep[EVE;][]{woods} on board the \textit{Solar Dynamic Observatory} (SDO). We also utilized X-ray observations provided by the Fermi Gamma-Ray Burst Monitor \citep[GBM;][]{meegan}, focusing on emissions in the 4--100~keV range. Observations at 17 and 34 GHz from the Nobeyama Radioheliograph \citep[NoRH;][]{nak_norh} are additionally investigated.

The excellent signal-to-noise ratio of the GOES-15 instrument allow us to study the derivative of a solar flare time series in detail, revealing a wealth of pulsations and fine structure \citep{simoes}. It is now certain that this type of variability is real and of solar origin based on comparisons with simultaneous observations of different events made with both GOES 13 and 15\footnote{RHESSI Nugget \#262 \textit{Fine Structure in Flare Soft X-ray Light Curves} Dennis, B.R \& Tolbert, K.}, and with comparison with other instruments such as LYRA and ESP \citep{dolla}.

Figure \ref{fig:one}(a) gives an overview of the lightcurves under investigation. The impulsive nature of the flare is displayed by the non-thermal emission such as Fermi 25--100~keV and NoRH 17 and 34~GHz. This impulsive behaviour begins at 01:55:00~UT and continues with 8 distinctive peaks of growing intensity.  The thermal emission seems not to show an impulsive phase signature in Figure \ref{fig:one}(a), with the rise and decay profiles looking relatively smooth with each channel peaking at different times depending on their temperature response. However, Figure \ref{fig:one}(b) shows the time derivative of the SXR channels. The fine-scale structure of the SXR emission is visually evident with similar structure to the HXR lightcurves in Figure \ref{fig:one}(a).

The non-thermal HXR and microwave pulsations cease at around 02:00:30~UT. After this time the correlation between GOES, LYRA Zr and ESP 0.1--7~nm become less evident but pulsations continue into the decay phase in both GOES channels. 
To investigate this further, the flare was broken into two phases - the impulsive phase: 01:55:00--02:00:30~UT and the gradual phase: 02:00:30--02:20:00~UT.
%and characteristics of these intervals were studied with cross-correlation and wavelet analysis techniques.

\section{Results}

\subsection{Impulsive Phase QPP}
The impulsive nature of this flare is seen from the onset of pulsations at 01:55:00~UT until 02:00:30~UT (marked within the red vertical lines in Figure \ref{fig:one}).  This `bursty' QPP regime is seen clearly in the non-thermal emissions with large modulation depths in the lightcurves, up to 80\% in the 50-100 keV. The HXR spectra displays a soft-hard-soft evolution of these modulated individual peaks. The modulation depth is calculated as the ratio of the amplitude of the pulsations to the to overall trend of the lightcurve. The maximum modulation depth is given for all channels in Table \ref{tab:1}. Each peak of a non-thermal pulsation is numbered in Figure \ref{fig:one} for comparison with the SXR derivatives. Notably some pulsations in the SXR derivative appear to peak before the pulsations in the HXR, such as peak 2, 3, 6 and 7. This seems inconsistent with the Neupert effect \citep{neupert,dennis_zarro}, in which we would expect the SXR derivatives to peak simultaneously with the HXR if we are to believe that the same electrons that produce the HXR also heat the plasma that produces the SXR.

In order to compare the short time-scale variability, the gradual varying trend of the lightcurves was removed by subtracting a smoothed time-series from the original time-series. The smoothed time-series was calculated by taking a boxcar average of the original time-series using a full width window of 30~s. This timescale window was chosen to highlight small scale fluctuations. For the thermal channels we used the original SXR lightcurves, rather than the derivatives. Subsampling to the longest cadence instrument (GOES), cross-correlation coefficients of each waveband with respect to GOES 1--8~\AA\ are presented in Table \ref{tab:1}. We find that on these short time scales, there is minimal delay between all wavebands.

\begin{figure}
\begin{center}
 \includegraphics[width=1.0\linewidth]{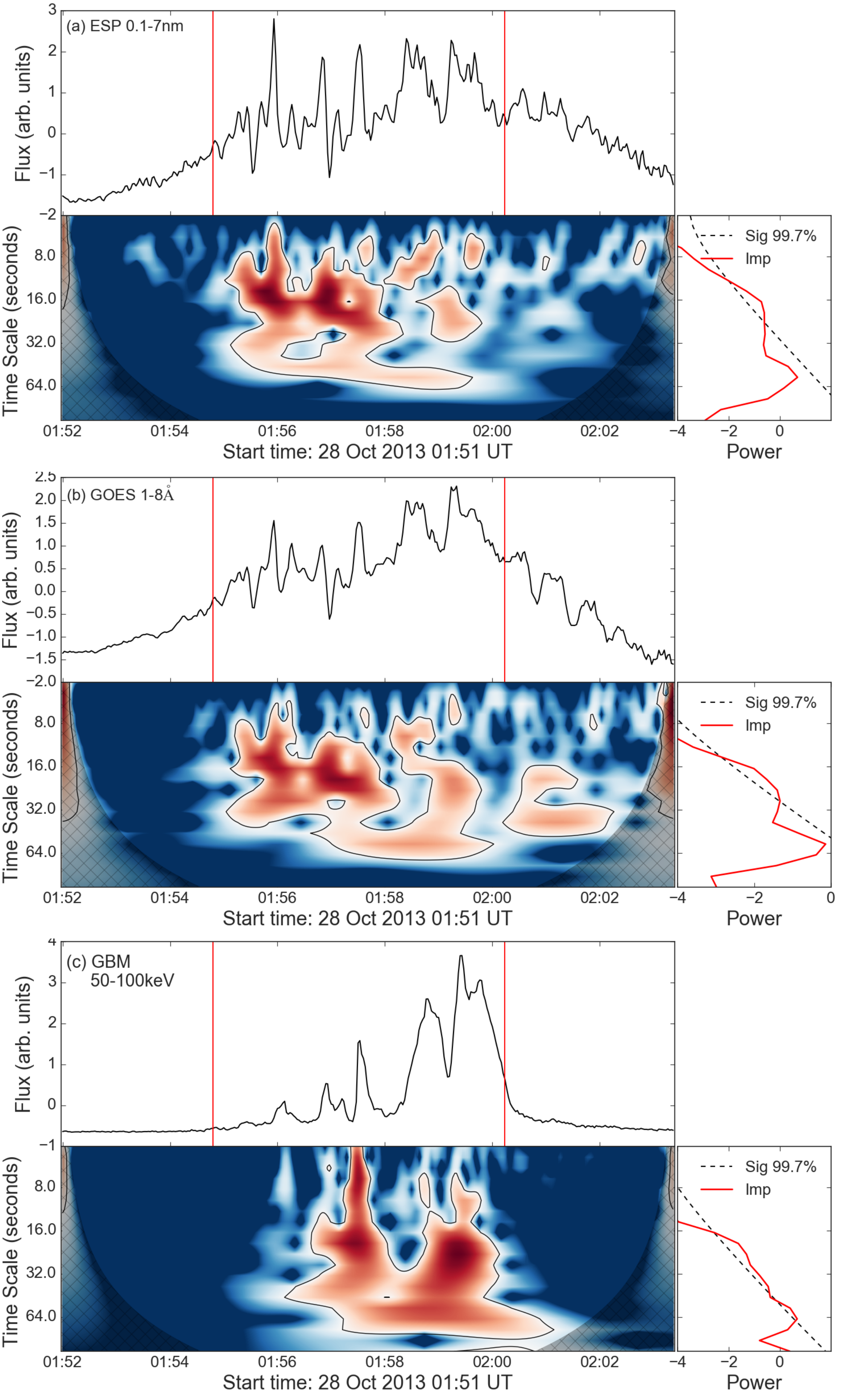}
\caption{Wavelet analysis of different channels. (a) ESP 0.1--7~nm derivative, (b) GOES 1--8~\AA\ derivative and (c) GBM 50--100~keV.}
\label{wave}
\end{center}
\end{figure} 

\begin{table*}
  \centering
\begin{tabular}{c c c c c c c c c c c}
\hline

Instrument  & Corr.Coef & Modulation (\%) & Power-law Index & Significant Timescale Range(s) \\

\hline
GOES 1--8~\AA &   1 & 0.9  & -2.3 & 14-27~s \\
GOES 0.5--4~\AA & 0.92 & 1.2 & -2.3 & 17-29~s   \\
LYRA Zr & 0.85 & 0.7 & -2.2 &13-24~s  \\
ESP 0.1--7~nm & 0.90 & 1.2 & -2.3 &   12-25~s\\
GBM 4--15~keV & 0.81 & 2.1  &-2.2 & 18-32~s \\
GBM 15--25~keV & 0.78 & 6.5& -2.1 &14-33~s \\
GBM 25--50~keV & 0.64  & 51.2 & -2.3 &16-26, 34-64~s\\
GBM 50--100~keV & 0.56  & 80.1 & -2.3 &17-40, 49-68~s\\
NoRH 17~GHz & 0.51 & 35.6& -2.3 &15-27, 48-70~s\\
NoRH 34~GHz &  0.48 & 16.2 & -2.3 &  17-30, 54-56~s \\

\hline
\end{tabular}
  \caption{Summary of characteristics of pulsations across multiple wavelengths during the impulsive phase. The cross correlation coefficients are shown compared to GOES 1-8~\AA. The `significant timescale'}  column gives the range of timescales for which the summed power exceeds the 99.7\% significance level above the power-law background model during impulsive phase.
  
  \label{tab:1}
\end{table*}

To search for characteristic timescales during the impulsive phase, wavelet analysis was employed using a Morlet wavelet. It was recently pointed out that the Fourier power spectra of many flare time series tends to approximate a power-law with negative index that tails off to a constant at higher frequencies, and that this must be taken into account when searching for periodicity. \citet{gruber} and \citet{inglis_red} demonstrated the dangers of assuming a flat power spectrum when studying oscillatory signatures in flare time series. They found that what may look like an oscillation is often not statistically distinguishable from a fluctuation in a power-law-like power spectrum. Following this, we test for the significance of enhanced power assuming a background power-law spectrum, estimated for each individual time series using an auto-regressive AR(1) model as described in \citet{tor}. The power-law slope that resulted from this is given in Table \ref{tab:1} for each time series.

 Figure \ref{wave} shows the results of the wavelet analysis for the different time series spanning thermal and non-thermal emission. Figure \ref{wave}(a), (b) and (c) show wavelet analysis of ESP 0.1--7~nm, GOES~1--8~\AA\, and GBM 50--100~keV, respectively. Each shows the time series under investigation, the wavelet power spectrum and the global time-averaged wavelet spectrum. During the impulsive phase, enhanced power is broadband with no narrow feature of a single timescale present. The dotted line in the global power spectrum is at the 99.7\% significance level above the power-law background model. We find that in all channels our time averaged global power spectrum shows a characteristic timescale at $\sim$20~s. A second peak is seen at $\sim$55~s but only reaches above the 99.7\% significance level in the non-thermal channels. The range of characteristic timescales for each channel which are above the significance level are listed in Table \ref{tab:1}.

\subsection{Decay phase QPP}
\noindent The gradual phase of the flare has a different nature to the highly correlated `bursty' qualities of the impulsive phase. Modulated emission features in HXR and microwave cease at around 02:00:30 UT. However, pulsations in the GOES time derivative persist well into the decay phase even though the non-thermal pulsations have ceased. Some of the extended variability is seen in ESP 0.1-7~nm, but is much less pronounced and almost nothing can be detected above noise in LYRA Zr. This could be due to the precision of the instruments, or due to the fact that both these channels are also sensitive to lower temperature plasma. This raises the question of what the temperature distribution of the pulsating plasma is, and requires further investigation outside the scope of this letter.

\begin{figure}
\begin{center}

  \includegraphics[width=1.0\linewidth]{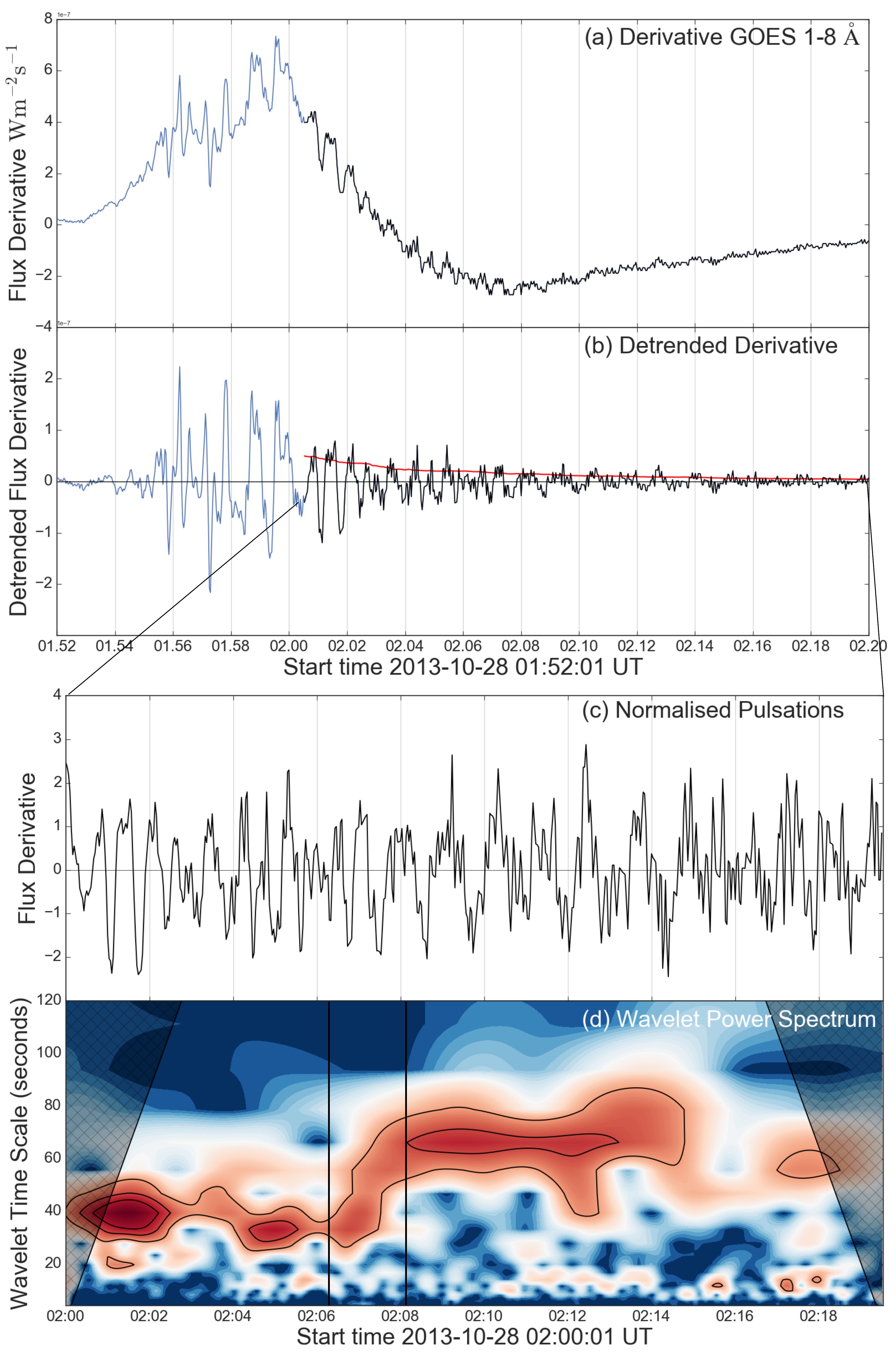}
\caption{Decay phase QPP observed by GOES 1-8~\AA. The black color indicates the decay phase section of the lightcurve we are interested in. (a) Derivative of the GOES 1--8 \AA\ lightcurve. (b) Detrended GOES derivative. (c) Normalized amplitude pulsations. (d) Wavelet power spectrum of (c). The vertical black lines at 02:06:30 and 02:08:30~UT highlight the region in time in which the characteristic timescale increases from $\sim$40~s to $\sim$70~s.}
\label{grad_deriv}
\end{center}
\end{figure}

In order to investigate the extent of the pulsations in the decay phase of this flare, we looked at both GOES channels from 02:00:30 UT to 02:20:00 UT. Figure \ref{grad_deriv} (a) and (b) shows the time derivative and detrended time derivative lightcurves of the GOES 1-8~\AA\ channel. The extended nature of the pulsations is clearly demonstrated. The fine structure features appear to have a less chaotic nature compared to the impulsive phase with the pulsations displaying a damped signature.

We again employ wavelet techniques to investigate the timescales of these pulsations during this phase. Wavelet power is dependent on the amplitude of the pulsations, and so due to the damping nature observed here, we first we first normalize the pulsations to have a constant amplitude. The normalization is done by dividing the time series in Figure \ref{grad_deriv}(b) by an envelope (marked by the red curve), calculated as the absolute value of the lightcurve smoothed with a boxcar window of $\sim$200~s.
Figure \ref{grad_deriv}(c) show these normalized amplitude pulsations during the decay phase. A wavelet power spectrum of the variations is shown in \ref{grad_deriv}(d). Interestingly, enhanced power is seen at a timescale of $\sim$40~s just after the impulsive phase at 02:00:30~UT which then changes to $\sim$70~s between 02:06:30 to 02:08:30~UT. The timescale of $\sim$70~s then stays constant until approximately 02:15~UT, when the signal-to-noise level becomes comparable to the amplitude of the pulsations.

\begin{figure*}
\begin{center}

 \includegraphics[width=1.0\linewidth]{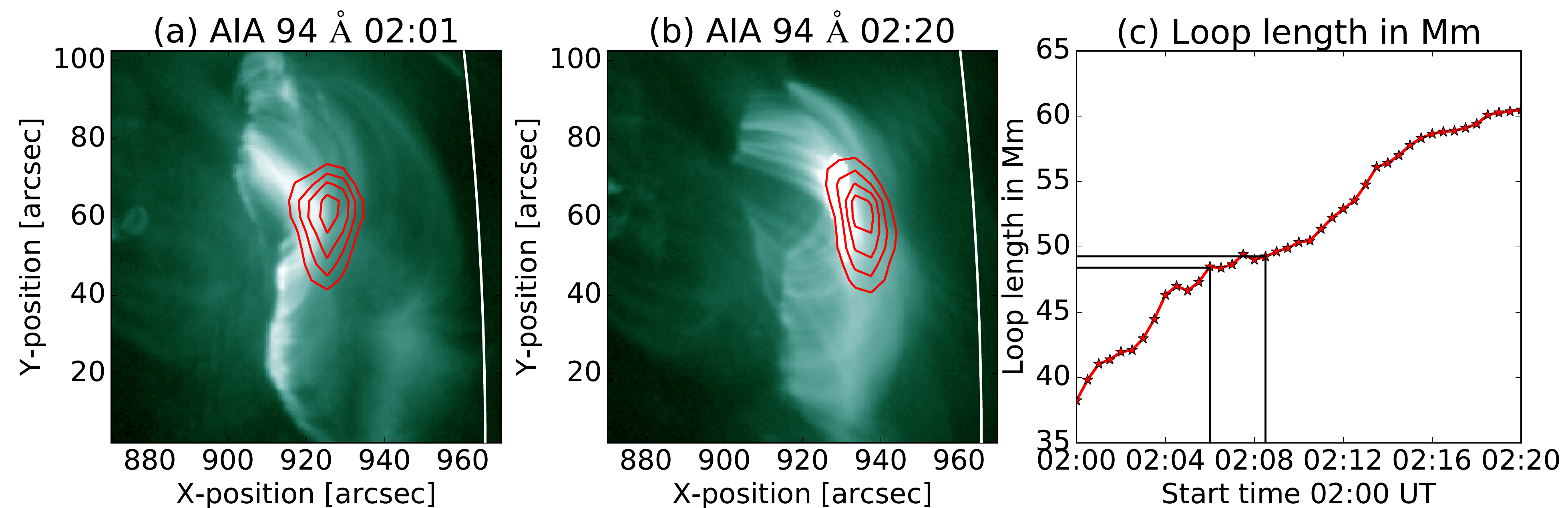}
\caption{(a) and (b): RHESSI contours of the 6--12~keV energy range using the CLEAN reconstruction algorithm overlaid on AIA 94~\AA\ images during the decay phase of the flare at 02:01~UT, and 02:20~UT respectively. The contours are at 30, 50, 70 and 90\% of peak value. (c) The loop length increase of the flare during the decay phase. The region noted within the black vertical lines is the time when the timescale increases from $\sim$40~s to $\sim$70~s.}
\label{test}
\end{center}
\end{figure*}

This increase in timescale of the pulsations is likely to be connected with longer loop lengths at later stages of the flare. Figure \ref{test}(a,b) shows the contours of the SXR source of RHESSI 6--12~keV overlaid on AIA 94~\AA\ images. We found that the SXR source height increases as the flare progresses, presumably a signature of continued magnetic reconnection resulting in newly formed hot loops at higher altitudes. Assuming semicircular, vertical loops seen in the plane of sky we estimate the loop length. This is done by locating the centroid of the SXR source imaged with RHESSI and approximating footpoints from AIA images. We then divide this observed length by $cos(20^{\circ})$ to account for projection affects from being $\sim20^{\circ}$ inside the limb. We find that the loop length increased from approximately 38~Mm at 02:00~UT to 59~Mm at 02:20~UT. This increase is plotted as a function of time in Figure \ref{test}(c). Notably during the increase in timescale from 02:06:30 to 02:08:30~UT, the loop length only increases by 2~Mm, which is small compared to the large increase in timescale. The loop length  then continues to grow even when the timescale of the pulsations is constant at 70~s until 02:15~UT.

\section{Discussion and Conclusion}

We have detected and analysed pulsations observed at multiple wavelengths during the X1.0 flare of 2013 October 28. Throughout the impulsive phase of the flare, highly correlated common features are observed at HXR, SXR, and microwave wavelengths with minimal time delay between peaks. Wavelet analysis of this impulsive interval shows broadband features in the wavelet power spectrum, with similar enhanced power in all channels. Characteristic peaks in the global spectrum at $\sim$20~s are detected in all wavebands with enhanced power also seen at around $\sim$55~s but only with significance above 99.7\% in the non-thermal emissions. These characteristic timescales are consistent with previous QPP investigations of different events. \citep[e.g.][]{Kupriyanova_2010, simoes, inglis_new}

 After the highly correlated impulsive phase, we find that emission in the non-thermal channels is no longer present. However, distinct pulsations in the high temperature plasma observed by GOES persist into the decay phase. The timescale of the pulsations are seen to increase from $\sim$40~s\ at the end of the impulsive phase at 02:00~UT to $\sim$70~s at 02:15~UT. These thermal pulsations could be a manifestation of continuing weak particle acceleration  \citep[e.g.][]{maccombie}, or some other heating mechanism that persists into the decay phase of the flare. This would support the idea that continuous heating is required to describe the decay times observed in many flares which are longer than the estimated conduction and radiation cooling timescales \citep{ryan, cargill}.

But what controls the timescale of the observed pulsations? The timescale of the QPP is consistent with expected characteristic timescales of MHD modes in the corona \citep{mcewan, pascoe07, pascoe09, macnamara}. However, given the complex evolution of geometrical plasma structures occurring during the impulsive phase, the identification of the specific MHD modes of oscillation producing the QPP is unlikely. The large modulation depths of the non-thermal emission during this time suggest that the pulsations are a result of episodic reconnection. The timescale would then be determined by either dynamic or periodic variations of the magnetic reconnection process such as multi-island reconnection in coronal current sheets \citep{drake06, guidoni}.

During the decay phase however, the increase in timescale and small amplitude of the thermal SXR pulsations is consistent with MHD processes within the flaring site. Recent studies have attributed persistent SXR QPP to that of compressive MHD modes such as the global fast sausage mode \citep{tian} and vertical kink mode \citep{dennis}. Both of these compressive modes would result in the observed pulsations in the decay phase of this flare. The damped nature of the decay phase pulsations observed in Figure \ref{grad_deriv}(b) suggest that they are a result of the global sausage mode in the leaky regime. In the leaky regime, sausage modes are subject to damping and show a decaying oscillatory behaviour. The period of the sausage mode is determined by the ratio of the wavelength (twice the loop length $L$) to the external Alfv\'en speed: $P = 2L/V_{Ae}$ \citep{pascoe07}. This dependence decreases in the leaky regime, and in the long-wavelength limit, the period becomes independent of the length of the oscillating loop and is determined by the transverse travel time across the loop; $P^{leaky} \approx \pi a/V_{Ai}$, where $a$ is the loop width and $V_{Ai}$ is the internal Alfv\'en speed \citep{nak_12}. This may explain why the characteristic timescale stays constant at $\sim$70~s even when the loop length keeps increasing. Taking the period to be 70~s and $a$ in the range of $\sim$ 2-10'', this interpretation yields an estimation of $V_{Ai} \approx 65 - 350kms^{-1}$. This is considerably low for coronal loop Alfv\'en speeds \citep[e.g.][]{asch_vel}, suggesting that the loops are not sufficiently long enough to be in the long-wavelength limit and so this interpretation cannot fully explain our observations. However, given the low signal-to-noise in the late stage of the flare, it is difficult to determine the true nature of these pulsations.

We cannot conclusively determine what mechanism generates these pulsations. However, the analysis of the SXR fine structure across multiple channels and its relation with other energies provides a new diagnostic tool. When correlations between different energies are high, especially SXR and HXR (as with most peaks in the impulsive phase of this flare), we can argue that the SXR emitting plasma is heated by electron beams at that time. When pulsations in SXR are seen to occur before HXR emission, and persist late into the decay phase after the HXR emission has stopped, some other heating mechanism is taking place. Thus, detailed comparisons between thermal and non-thermal structures
can potentially help us understand the distribution of the different types of heating taking place during a flare. Additionally, it has now become apparent that fine structure pulsations observed in the SXR derivative are a common \citep{simoes}, and maybe even an intrinsic feature of flaring emission. Hence, further investigation into the structure evident in SXR lightcurves and multi-wavelength comparisons will provide better insight into the origins of the QPP phenomena.

\acknowledgments
This work has been supported by Enterprise Partnership Scheme studentship from the Irish Research Council (IRC) between Trinity College Dublin and Adnet System Inc.  D.\ Ryan thanks the Solar-Terrestrial Centre of Excellence and the SIDC Data Exploitation and the NASA Postdoctoral Program — administered by the Universities Space Research Association — for their financial support. The support of the PROBA2 Guest Investigator Program provided opportunity to collaborate with the PROBA2 team at the Royal Observatory Belgium. This research has made use of SunPy, an open-source and free community-developed solar data analysis package written in Python \citep{mumford}. We also acknowledge the anonymous referee whose comments helped to improve the paper.

\vspace{5mm}
\facilities{GOES, Fermi (GBM), NoRH, PROBA2 (LYRA), RHESSI, SDO (EVE, AIA)}



\end{document}